\documentclass{ws-procs9x6}

\begin{document}

\title{Transverse target-spin asymmetry associated with DVCS on the proton and a resulting model-dependent constraint on the total angular momentum of quarks in the nucleon}

\author{Zhenyu Ye\\
        on behalf of the HERMES collaboration}

\address{DESY, 22607 Hamburg, Germany\\
         E-mail: yezhenyu@mail.desy.de}

\maketitle

\abstracts{
Results are reported on the transverse target-spin asymmetry associated with deeply virtual Compton scattering on the proton, extracted from the data accumulated by the HERMES experiment in the years 2002-2004. By comparing the HERMES results and theoretical predications based on a phenomenological model of generalized parton distributions, a model-dependent constraint on the total angular momentum carried by quarks in the nucleon is obtained.}

\section{Introduction}
Deeply virtual Compton scattering (DVCS) is an exclusive process in which a virtual photon (emitted by an incoming lepton) is absorbed and a real photon is produced by a single parton in the nucleon, the recoiling nucleon being in its ground state. DVCS is one of the theoretically cleanest processes to access generalized parton distributions (GPDs) which provide a detailed description of the nucleon structure. Great interest in them has arisen after it was realized that the total angular momentum carried by quarks in the nucleon, $J_q$, may be obtained\cite{Ji97} from the GPDs $H_q$ and $E_q$ for the quark species $q$.

The transverse target-spin asymmetry (TTSA) associated with DVCS on the proton, measurable using an unpolarized (U) lepton beam and a transversely (T) polarized hydrogen target, is defined as\cite{Elli05},
\begin{equation}
A_{UT}(\phi,\phi_S)=\frac
{d\sigma(\phi,\phi_S) - d\sigma(\phi,\phi_S+\pi)}
{d\sigma(\phi,\phi_S) + d\sigma(\phi,\phi_S+\pi)},\label{eqn:ttsa}
\end{equation}
where $\phi$ denotes the azimuthal angle between the plane containing the incoming and outgoing lepton momenta and the plane correspondingly defined by the virtual and the real photon, and $\phi_S$ the one of the target polarization vector with respect to the lepton plane. Two azimuthal amplitudes of the TTSA, $A_{UT}^{\sin(\phi-\phi_S)\cos{\phi}}$ and $A_{UT}^{\cos(\phi-\phi_S)\sin{\phi}}$, appear to leading order in $\alpha_S$ and $1/Q$. They can be approximated as:
\begin{eqnarray}
A_{UT}^{\sin{(\phi-\phi_S)}\cos{\phi}} &\propto& Im\left[F_2\mathcal{H}-F_1\mathcal{E}\right], \nonumber \\
A_{UT}^{\cos{(\phi-\phi_S)}\sin{\phi}} &\propto& Im\left[F_2\mathcal{\widetilde{H}}-F_1\xi\mathcal{\widetilde{E}}\right].\label{eqn:ref1}
\end{eqnarray}
Here ${\mathcal{H}}$, ${\mathcal{E}}$, ${\mathcal{\widetilde{H}}}$ and ${\mathcal{\widetilde{E}}}$ denote convolutions of the respective GPDs $H$, $E$, $\widetilde{H}$ and $\widetilde{E}$ with hard scattering kernels, $F_1$ and $F_2$ are the Dirac and Pauli form factors of the proton, respectively. 

A parametrization for GPDs has been proposed by Goeke {\it et al.}\cite{Goeke01} where the GPD $E$ is modelled using $J_u$ and $J_d$ as free parameters. For this GPD model it has been found that the TTSA amplitude $A_{UT}^{\sin(\phi-\phi_S)\cos{\phi}}$ is sensitive to $J_u$ (and $J_d$)\footnote{In electroproduction on the proton $u$ and $d$ quark contributions have relative weight $4:1$ due to the squared quark charges.}, and insensitive to the other parameters\cite{Elli05}. In this talk we will report the first results obtained at HERMES on the TTSA associated with DVCS on the proton, and on a model-dependent constraint on $J_u$ vs $J_d$ obtained by comparing the HERMES results and theoretical predictions based on the above mentioned GPD model.

\section{The HERMES Experiment}
At HERMES the nucleon spin structure is studied using the 27.6 GeV electron (or positron) beam at HERA and internal polarized gaseous targets. A forward spectrometer instrumented with tracking chambers provides momentum and angular measurements for charged particles. Lepton-hadron separation is achieved by a transition-radiation detector, a pre-shower counter and an electromagnetic calorimeter, which also detects photons. Not all the hadrons in the final state are detected by the forward spectrometer -- the recoiling proton in the DVCS process typically travels perpendicularly to the beam direction and hence escapes the detector acceptance. The exclusivity of the selected events, which contain an identified scattered lepton and a produced real photon, is maintained by a missing-mass cut $-(1.5)^2<M_x^2<(1.7)^2$ GeV$^2$. Monte Carlo studies have shown that the non-exclusive contributions to the selected data sample originates mainly from semi-inclusive $\pi^0$ production amounting to approximately $5\%$. The contribution from the associated exclusive reaction, where the nucleon is excited to a resonant state, amounts to approximately $11\%$. 

\section{Results}
HERMES was taking data with a transversely polarized hydrogen target in the years 2002-2005. The measured kinematic dependence of the TTSA amplitudes has been reported elsewhere\cite{Ye05}. Here we report the integrated result $\langle A_{UT}^{\sin{(\phi-\phi_S)}\cos{\phi}}\rangle=-0.149$ $\pm$ 0.058(stat.) $\pm$ 0.033(syst.), extracted from the 2002-2004 data at the average kinematics $\langle -t\rangle$=0.12 GeV$^2$, $\langle x_B\rangle$=0.095, $\langle Q^2\rangle$=2.5 GeV$^2$. Corrections for semi-inclusive background and smearing have been applied. The main contributions to the systematic uncertainty are those from the determination of the target polarization, in the background correction, and due to acceptance effects. 

In order to constrain $J_u$ and $J_d$, the reduced $\chi^2$ value, defined as
\begin{equation}
\Delta\chi^2=\chi^2-\chi^2_{minimum}=\left[A^{exp}-A^{VGG}(J_u,J_d)\right]^2/\left[\delta A_{stat}^2+\delta A_{syst}^2\right],\label{eqn:chi2}
\end{equation}
is calculated for different values of $J_u$ and $J_d$. Here $A^{exp}$ denotes the measured (integrated) TTSA amplitude, $\delta A_{stat}$ ($\delta A_{syst}$) its statistical (systematic) uncertainty, and $A^{VGG}$ is the value calculated at the average kinematics by a code\cite{VGG} based on the GPD model proposed by Goeke {\it et al.}\cite{Goeke01}. As theoretical predictions on $A_{UT}^{\cos{(\phi-\phi_S)}\sin{\phi}}$ show minor changes\cite{Elli05} with variations in $J_u$ and $J_d$, only the contribution from $A_{UT}^{\sin{(\phi-\phi_S)}\cos{\phi}}$ to the reduced $\chi^2$ value is included in Eq.~(\ref{eqn:chi2}). 

\begin{figure}[t]
\centerline{\epsfxsize=0.8\columnwidth\epsfbox{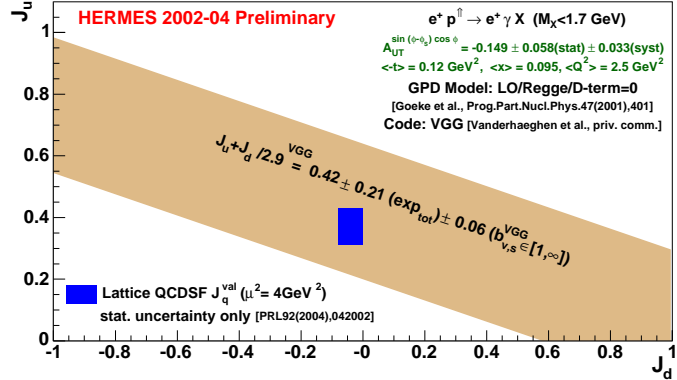}}
\caption{Model-dependent constraint on $u$-quark total angular momentum $J_u$ vs $d$-quark total angular momentum $J_d$, obtained by comparing the experimental result and theoretical predictions on the TTSA amplitude $A_{UT}^{\sin{(\phi-\phi_S)}\cos{\phi}}$. Also shown is a Lattice result from the QCDSF collaboration, obtained at the scale $\mu^2=4$ GeV$^2$ for valence quark contributions only.\label{fig:constraint1}}
\end{figure}

\begin{figure}[t]
\centerline{\epsfxsize=0.8\columnwidth\epsfbox{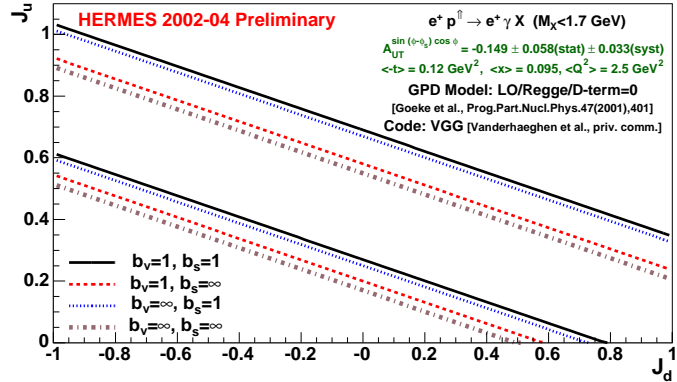}}
\caption{Model-dependent constraints on $u$-quark total angular momentum $J_u$ vs $d$-quark total angular momentum $J_d$ for different values of the profile parameter $b$.\label{fig:constraint2}}
\end{figure}

The area in the ($J_u$, $J_d$)-plane, in which the reduced $\chi^2$ value is not larger than one, is defined as the one-standard-deviation constraint on $J_u$ vs $J_d$. It is obtained to be $J_u+J_d/2.9=0.42\pm0.21\pm0.06$ (see Fig.~\ref{fig:constraint1}). The first uncertainty is due to the experimental uncertainty in the measured TTSA amplitude. The second one is a model uncertainty, obtained by varying from one to infinity the unknown profile parameter $b$ which controls the skewness dependence of GPDs\cite{Goeke01} (see Fig.~\ref{fig:constraint2}). The $t$-dependence of GPDs is modelled using the Regge ansatz\cite{Goeke01}. The impact of using it or its alternative -- the factorized ansatz -- on the theoretical predictions on the TTSA amplitudes has been found to be negligible\cite{Elli05}. The D-term contribution to the GPDs $H$ and $E$ is set to zero, as suggested by the HERMES results on the beam-charge asymmetry\cite{HERMES-BCA}. If the D-term were modelled according to the chiral quark soliton model\cite{Goeke01}, the resulting constraint is shifted to $J_u+J_d/2.9=0.53\pm0.21\pm0.06$. 

\end{document}